\documentstyle[11pt,paspconf,psfig]{article}

\input{epsf}

\begin{document}

\title{X-rays and Soft Gamma-rays from Seyferts, Radio Galaxies, and Black-Hole
Binaries}
\author{Andrzej A.  Zdziarski}
\affil{N.  Copernicus Astronomical Center, Bartycka 18, 00-716 Warsaw, Poland}

\begin{abstract}
Selected topics dealing with X-ray/soft $\gamma$-ray emission from accreting
black-hole sources are reviewed.  The shape of soft $\gamma$-ray spectra of
Seyferts observed by OSSE can be well modeled by thermal Comptonization.  A
very strong correlation between the X-ray slope and the strength of Compton
reflection in these sources can then be explained by thermal Comptonization in
vicinity of the reflecting cold media.  Similar correlation is seen for the
hard state of black-hole binaries, but the strength of Compton reflection at a
given X-ray slope is larger than in Seyferts.  This appears to be due to seed
photons (for Comptonization) emitted by the cold media having larger energies
in stellar-mass systems than in Seyferts.  Broad-line radio galaxies have hard
X-ray spectra with weak reflection, which reflection is similar to that in
Seyferts with hard X-ray spectra.  Two main models for the source geometry, a
hot inner disk surrounded by a cold outer disk, and a dynamic corona above a
cold disk, appear about equally viable based on the presently available data.
\end{abstract}

\section{Introduction} \label{s:intro}

The origin of X-rays and soft $\gamma$-rays (hereafter X$\gamma$) from
accreting black-hole sources has been a major problem of high-energy
astrophysics for many years. Currently, we can probably identify main radiative
processes taking place in those sources, but their geometry and dynamics remain
still unknown.

In this paper, I first present a short general discussion of various classes of
X$\gamma$ sources around accreting black holes.  This is followed by discussion
in more detail of selected specific topics:  soft $\gamma$-ray spectra of
Seyferts, Compton reflection, and X$\gamma$ emission from broad-line radio
galaxies (hereafter BLRGs), with emphasis on radiative processes giving rise to
observed spectra.  Finally, I discuss some constraints on geometry of the
sources.  Recent related reviews include Poutanen (1998) and Zdziarski et al.\
(1997, hereafter Z97).

\section{Overview} \label{s:over}

\begin{figure} 
\centerline{\psfig{file=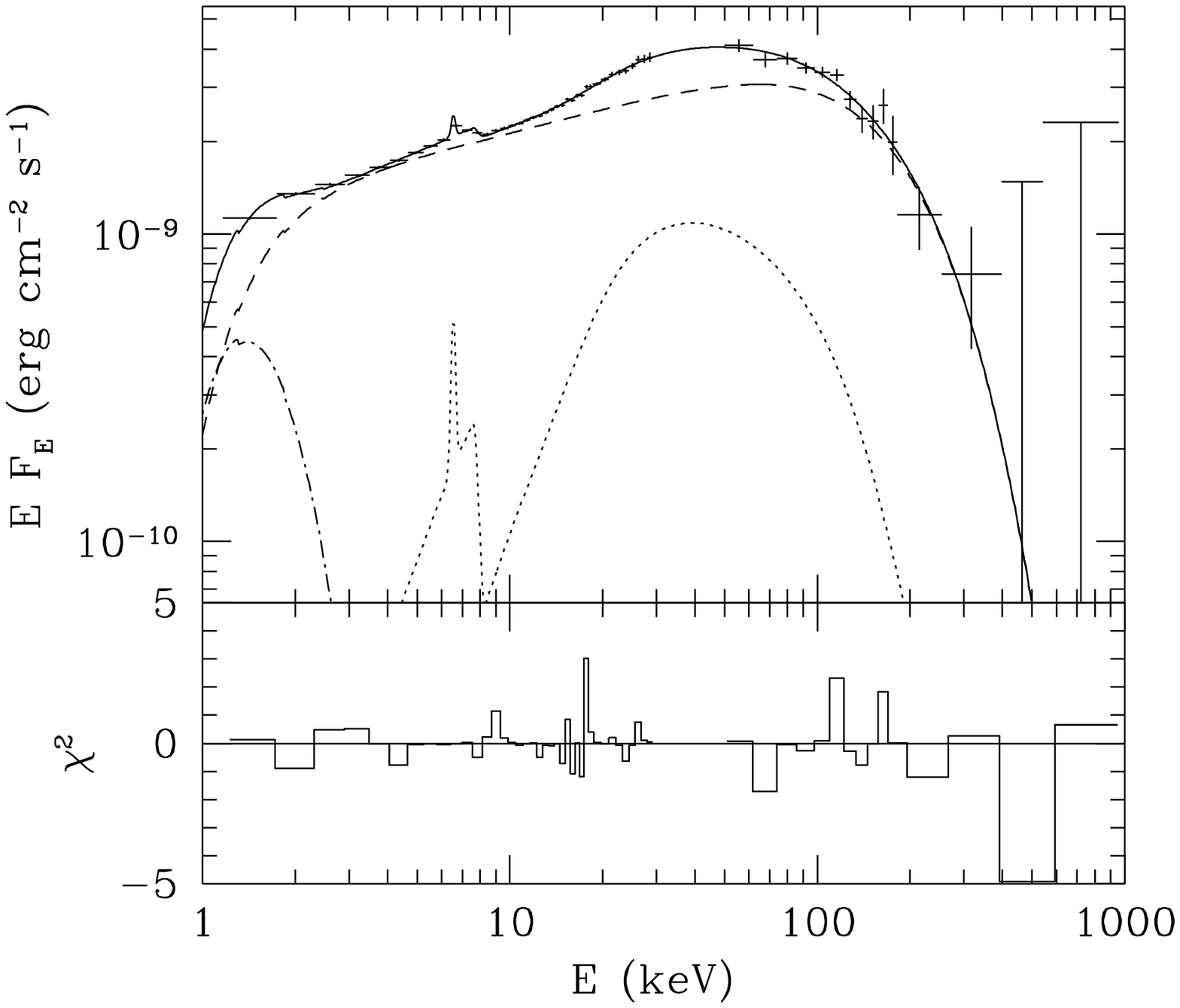,width=7cm,height=5.8cm}
\psfig{file=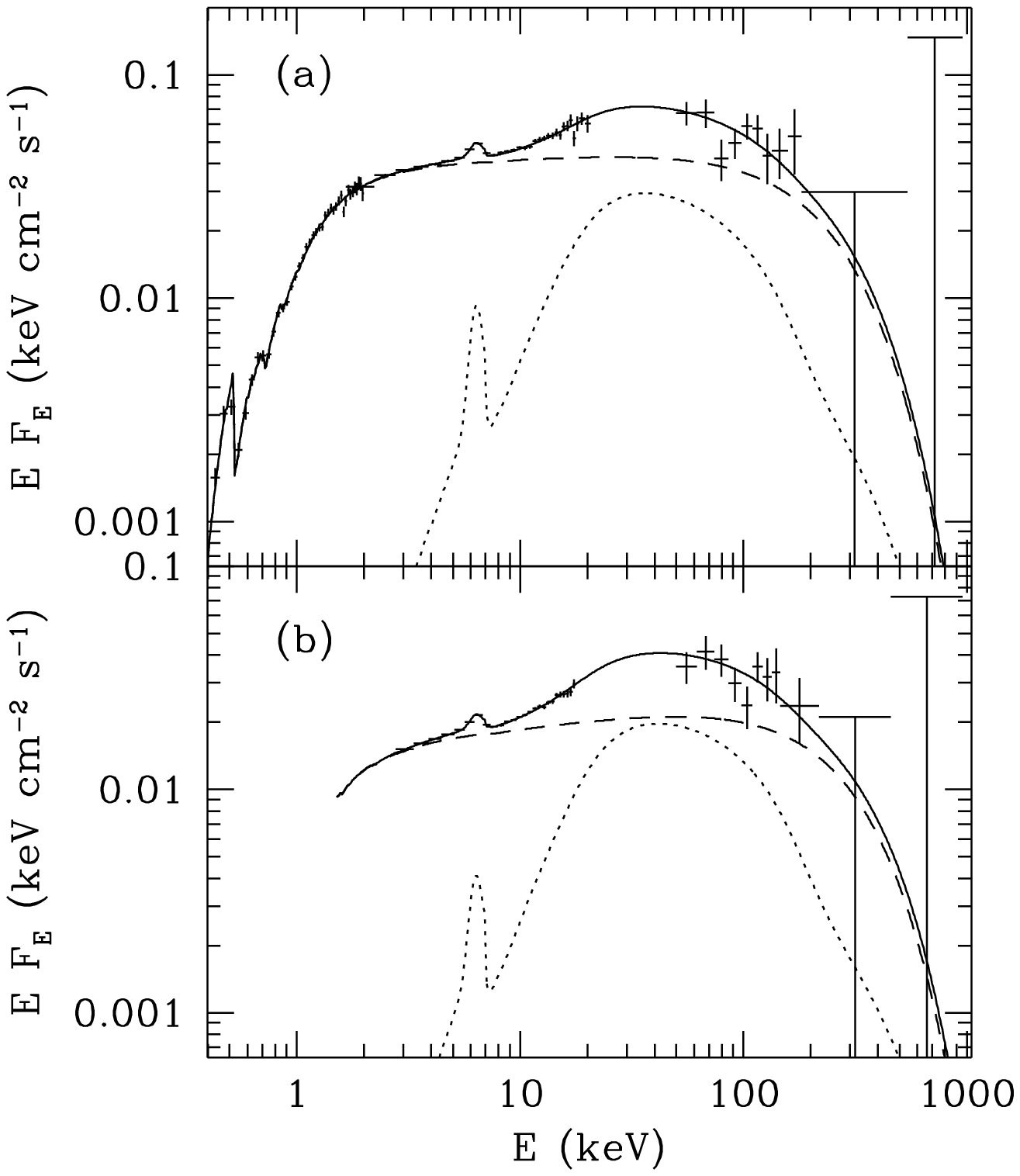,width=6cm,height=5.8cm}} 
\caption{\small The X$\gamma$ spectra
of GX 339--4 (a black-hole binary) in the hard state (left panel), of IC 4329A,
a Seyfert 1 (right upper panel), and of the average of 5 Seyfert 1s detected by
both {\it Ginga}\/ and OSSE (right lower panel).  The solid curves correspond
to models consisting of thermal Comptonization in a plasma with $kT\sim 10^2$
keV and $\tau\sim 1$ (dashes), a reprocessed component with a
Compton-reflection continuum and a fluorescent Fe K$\alpha$ line (dots), and,
in the case of GX 339--4, a soft blackbody emission (a dot-dashed curve).  }
\label{xgamma} \end{figure}

We can distinguish the following classes of X-ray sources powered by accretion
onto a black hole.  First, there are black-hole binaries, which have either
persistent or transient X-ray sources.  These sources are in one of two main
spectral states:  a soft (also called high) one or a hard (also called low)
one, and they often switch between the states.  Then radio-quiet (hereafter RQ)
Seyfert 1s are nearby AGNs without strong jets and observed without strong
obscuration.  X$\gamma$ spectra of Seyfert 1s are remarkably similar to
X$\gamma$ spectra of black-hole binaries in the hard state.  Seyfert 2s, most
likely, belong to the same intrinsic class as Seyfert 1s, but are oriented
close to edge-on, which results in strong obscuration both in the optical/UV
and X-ray ranges. Finally, radio galaxies form a rather enigmatic X-ray
class, with their X-ray properties somewhat similar to those of Seyferts.

A distinct class appears to be formed by Narrow-Line Seyfert 1s, whose X-ray
spectra are very soft, and within observational uncertainties, similar to those
of black-hole binaries in the soft state (Pounds, Done \& Osborne 1995).
However, very little is known about their hard X-ray and soft $\gamma$-ray
properties.

Hard X-ray spectra of basically all the considered classes of sources consist
of (intrinsic) power laws (with the photon number flux $\propto E^{-\Gamma}$
and typical spectral indices of $\Gamma\sim 1.5$--3) accompanied, in most
cases, by signatures of reprocessing by cold media:  Fe K$\alpha$ lines (e.g.\
George \& Fabian 1991) and Compton reflection (Lightman \& White 1988;
Magdziarz \& Zdziarski 1995).

The intrinsic hard X-ray power laws often break at $\sim 100$ keV.  The form of
the spectra in that range is known most accurately for black-hole binaries.
Their spectra in the hard state show high-energy breaks with a characteristic
curvature (Grove et al.\ 1998), which is well modeled by thermal
Comptonization in a plasma with a temperature, $kT\sim 100$ keV, and a Thomson
optical depth, $\tau\sim 1$ (Gierli\'nski et al.\ 1997; Zdziarski et al.\ 1998,
hereafter Z98).  A typical example of such a spectrum is that of of the hard
state of GX 339--4, a black hole candidate, shown in the left panel of Fig.\
\ref{xgamma} (Z98).

On the other hand, no distinct high-energy cutoffs are seen in the soft state
of black-hole binaries up to $\sim 1$ MeV (Grove et al.\ 1998; Phlips et al.\
1996; Tomsick et al.\ 1999).  This lack of cutoffs in soft $\gamma$-rays is
likely a signature of Compton scattering by non-thermal, relativistic,
electrons with a distribution close to a power law (Poutanen \& Coppi 1998;
Gierli\'nski et al.\ 1999).

Due to the relative weakness of their observed soft $\gamma$-ray fluxes, the
shape of high-energy cutoffs of AGNs is known much less accurately.  Still, NGC
4151, the RQ Seyfert brightest in soft $\gamma$-rays, has intrinsic X$\gamma$
spectra well modeled by thermal Comptonization in a plasma with $kT$ and
$\tau$ very similar to those of black-hole binaries (Zdziarski, Johnson \&
Magdziarz 1996; Johnson et al.\ 1997a).  Two examples of broad-band X$\gamma$
spectra of Seyfert 1s are shown in Fig.\ \ref{xgamma} (Z97).  The right top
panel shows a spectrum of the Seyfert 1 IC 4329A, and the right bottom panel
shows the average spectrum of 5 Seyfert 1s detected by both {\it Ginga}\/ and
{\it CGRO}/OSSE (Gondek et al.\ 1996), both fitted with a model consisting of
thermal Comptonization and Compton reflection. Also, the average spectra of all
RQ Seyfert 1s and 2s detected by OSSE show a similarly curved shape as the
spectrum of NGC 4151 (Z97; \S 3 below).

The shape of soft $\gamma$-ray spectra of radio galaxies is known even less
accurately.  On one hand, their X$\gamma$ spectra do show breaks at $\sim 100$
keV (Wo\'zniak et al.\ 1998, hereafter W98), but, on the other hand, the soft
$\gamma$-ray spectra above the break are compatible with a power-law form,
without evidence for further curvature.  The X$\gamma$ spectrum of the
brightest radio galaxy, Cen A, shows a similar break at $\sim 100$ keV with a
power law above this break continuing up to $> 10$ MeV, and with some flux
detected even at $\ga 100$ MeV (Steinle et al.\ 1998).  Such spectrum is
incompatible with thermal Comptonization, and its modeling seems to require
non-thermal processes similar to those operating in blazars.

In soft X-rays, all these classes of sources show additional components above
extrapolation of the hard X-ray power laws to lower energies.  In Seyferts and
radio galaxies, such soft X-ray excesses are common below $\sim 1$ keV.
Typically, the intrinsic soft X-ray spectra are power-law like and, when
extrapolated to the UV range, join onto the observed UV spectra, see, e.g.\
Magdziarz et al.\ (1998).

Black-hole binaries show additional soft X-ray components as well.  For
example, Cyg X-1 in the hard state shows a blackbody component with a
temperature of $\sim 0.15$ keV together with a soft power-law component
intersecting the hard X-ray power law at $\sim 3$ keV (Ebisawa et al.\ 1996).
In the soft state, the blackbody temperature is higher, e.g.\ $\sim 0.4$ keV in
the case of Cyg X-1, but an additional soft-excess component in addition to
both the blackbody and the hard X-ray power law is still present (Gierli\'nski
et al.\ 1999).

The interpretation of the soft power-law excesses present in both Seyferts and
both spectral states of black-hole binaries remains relatively unclear.
Plausibly, they are due to Comptonization by a thermal plasma with a small
Compton parameter [$y\equiv 4(kT/m_{\rm e}c^2)\max(\tau, \tau^2)\ll 1$] in addition
to Comptonization by another plasma forming the hard X-ray power law (e.g.\
Magdziarz et al.\ 1998; Gierli\'nski et al.\ 1999).  On the other hand, the
blackbody component in black-hole binaries and the blue-bump UV component in
Seyferts originate likely in an optically-thick accretion disk.

Finally, the observed spectra are attenuated by bound-free absorption due to
the Galactic column density towards the source.  Also, intrinsic absorption by
both neutral and ionized matter is commonly seen in AGNs.  On the other hand,
intrinsic absorption in black-hole binaries is typically transient (e.g.\ in
Cyg X-1, Ebisawa et al.\ 1996).

\section{Soft Gamma-ray Spectra of Seyferts} \label{s:soft}

As stated in \S 2, intrinsic broad-band X$\gamma$ spectra of Seyferts are well
modeled by thermal Comptonization with $kT\sim 10^2$ keV and $\tau\sim 1$.
These parameters are implied by the shape of high-energy cutoffs observed in
soft $\gamma$-rays by OSSE when combined with the spectra observed in X-rays.
Soft $\gamma$-ray spectra of Seyferts from OSSE are reviewed by Johnson et al.\
(1997b). The OSSE spectra are limited in photon statistics, and significant
constraints can thus be obtained for brightest objects and average spectra
only.

\begin{figure}
\centerline{\psfig{file=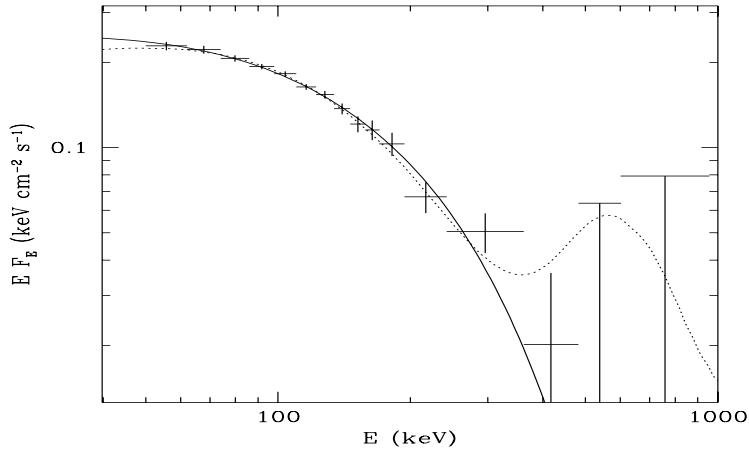,width=10cm,height=6cm}}
\caption{\small The average OSSE spectrum of NGC
4151.  The solid curve represents the best-fit thermal-Comptonization model.
The dotted curve corresponds to a hybrid thermal/non-thermal model with the
maximum allowed fraction ($\approx 15\%$) of the total power going into
electron acceleration (with the remainder used for electron heating).  }
\label{fig:4151} \end{figure}

The RQ Seyfert brightest in soft $\gamma$-rays, NGC 4151, was studied by
Zdziarski et al.\ (1996) and Johnson et al.\ (1997a).  They found that its
X$\gamma$ spectra are well modeled by thermal Comptonization in a plasma with
$kT\sim 60$ keV and $\tau\sim 1$.  These data can also be used to constrain a
possible presence of non-thermal electrons in the source.  A fraction of the
total power in the source going to acceleration of electrons to suprathermal
energies is $\la 15\%$ (and consistent with null), as illustrated in Fig.\
\ref{fig:4151}. Interestingly, the average OSSE spectrum of NGC 4151 is
virtually identical in shape to the 1991 September OSSE spectrum of GX 339--4,
a black-hole candidate (Z98; see Fig.\ \ref{xgamma} above).

We then consider co-added spectra of all RQ Seyfert 1s and of all Seyfert 2s
detected by OSSE through 1998 January (Zdziarski, Poutanen \& Johnson, in
preparation).  This represents a significant update of the average OSSE spectra
detected through 1995, which were presented in Z97.  The Seyfert-1 average
spectrum has been obtained by combining $1.4\times 10^7$ seconds of OSSE
exposure (scaled to a single OSSE detector) of 17 AGNs:  ESO 141--55, IC 4329A,
III Zw 2, MCG +8-11-11, MCG --2-58-22, MCG --6-30-15, Mkn 279, Mkn 509, Mkn
841, NGC 3227, NGC 3516, NGC 3783, NGC 526 A, NGC 5548, NGC 6814, NGC 7213, NGC
7469. The Seyfert-2 average spectrum has been obtained by combining $8.2\times
10^6$ seconds of exposure of 10 AGNs:  MCG --5-23-16, Mkn 3, NGC 1275, NGC
2110, NGC 4388, NGC 4507, NGC 4945, NGC 5506, NGC 7172, NGC 7582.

\begin{figure}
\centerline{\psfig{file=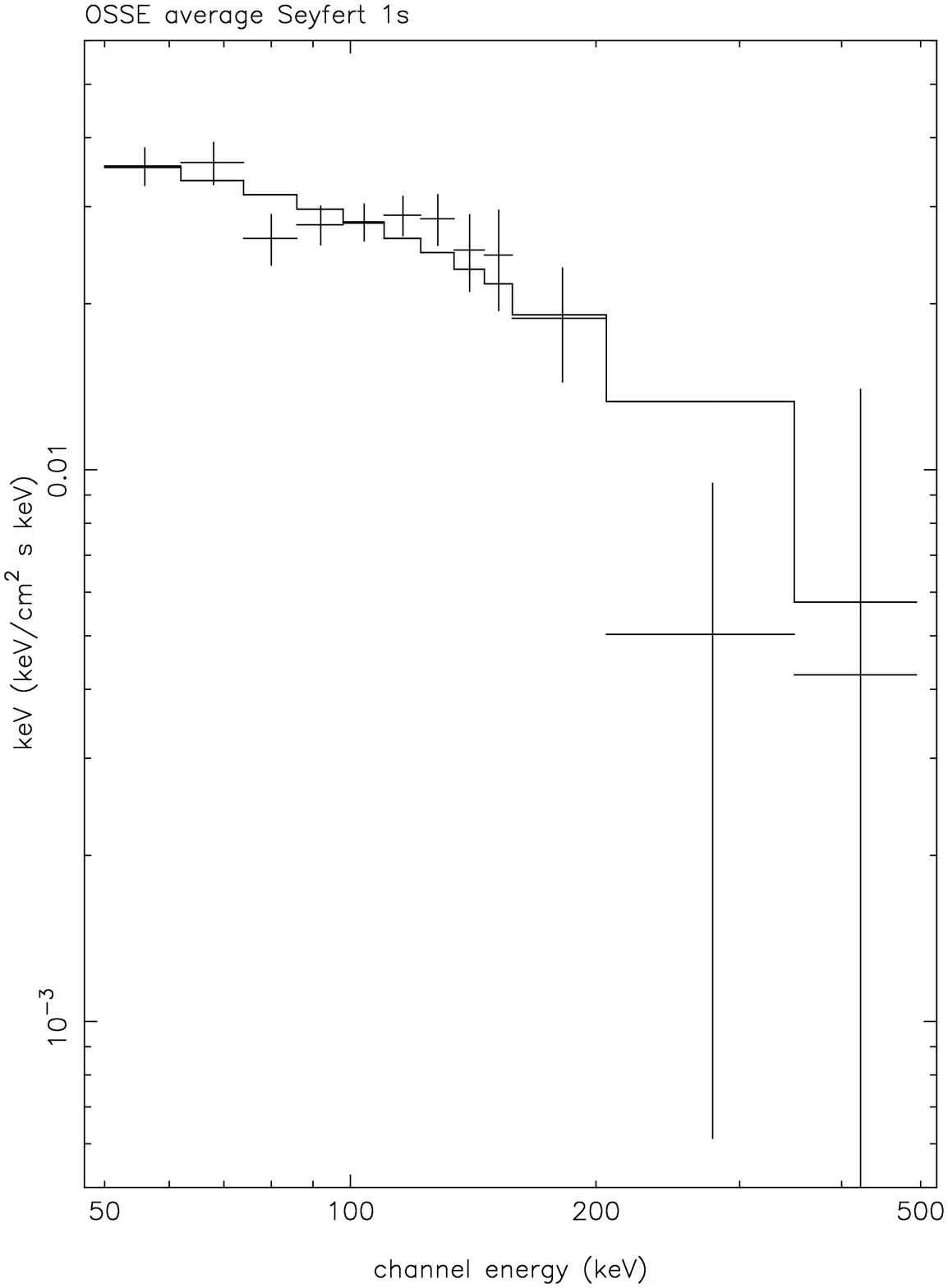,width=6cm,height=6cm}
\psfig{file=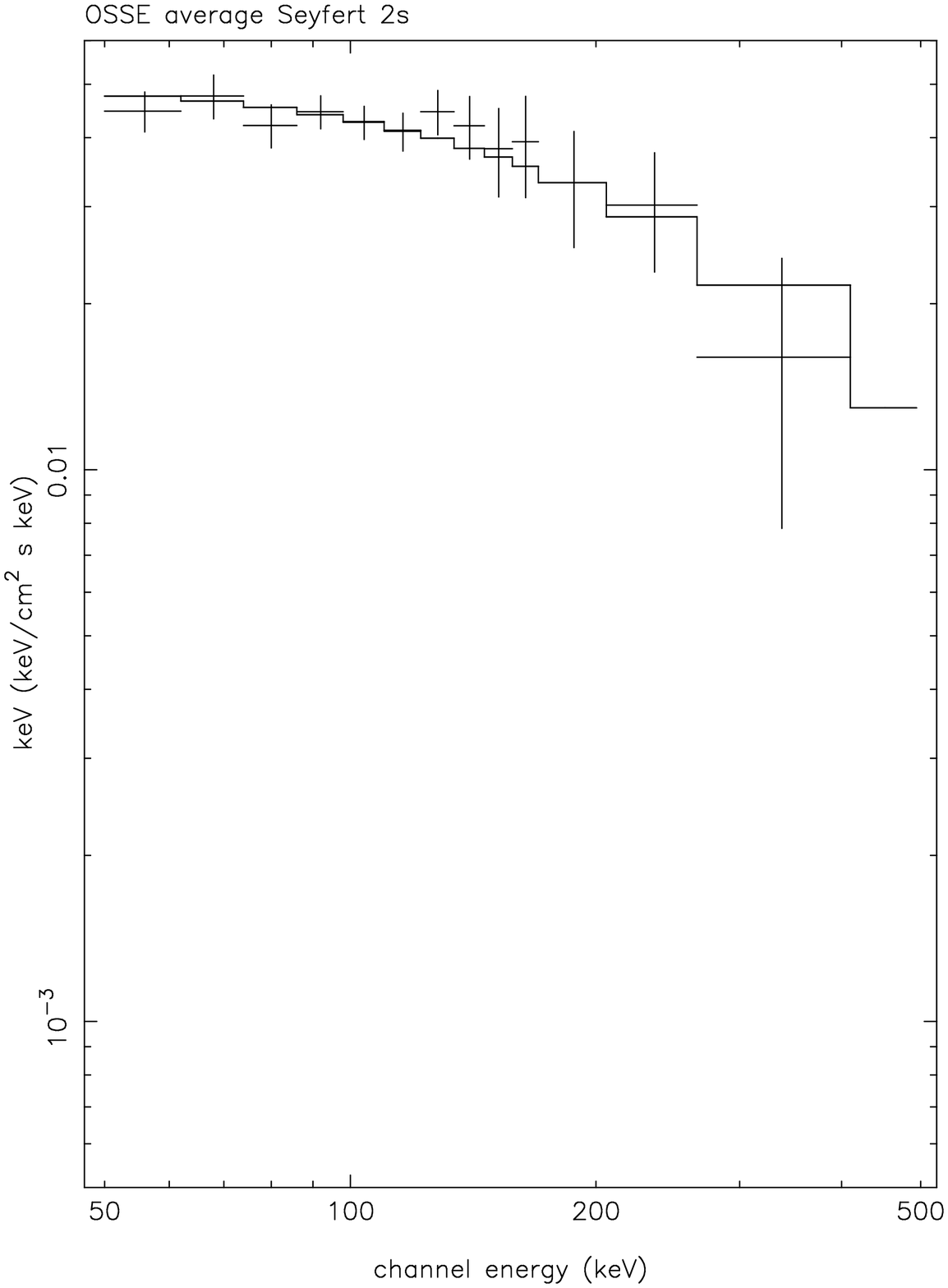,width=6cm,height=6cm}} 
\caption{\small The average OSSE spectra of Seyfert
1s and 2s, fitted by thermal Comptonization and Compton reflection.  }
\label{sy_osse} \end{figure}

We fit the spectra, shown in Fig.\ \ref{sy_osse}, in the 50--500 keV range.
First, we find a significant presence of high-energy cutoffs in the average
spectra.  Namely, the $\chi_\nu^2$ improves from 33/35 and 48/35 for power-law
fits to 25/34 and 34/34 for fits with an e-folded power law in the case of the
Seyfert-1 and Seyfert-2 average spectrum, respectively.  The statistical
significance of the presence of spectral curvature corresponds then to the
probability that the fit improvement were by chance of 0.002 and 0.0007 for
Seyfert 1s and 2s, respectively.  This result argues strongly that the spectra
of individual Seyferts are strongly cut off as well.  If, instead, their
spectra were simple power laws with a range of spectral indices, the sum
spectrum would have a concave shape rather than the observed convex one.

We then consider models with thermal Comptonization and Compton reflection.  We
find that the OSSE data alone cannot constrain the spectral index in the X-ray
range as well as the strength of reflection.  Therefore, we first use the
average X$\gamma$ spectrum of Seyfert 1s observed by both {\it Ginga}\/ and
OSSE of Gondek et al.\ (1996) (shown on the lower right panel of Fig.\
\ref{xgamma}). We fit that spectrum with a model of thermal Comptonization in a
hot plasma slab with sources of seed photons located in the midplane (Z98).  We
assume the inclination of a surrounding (reflecting) cold disk of $i=30^\circ$,
as expected for Seyfert 1s (e.g.\ Nandra et al.\ 1997).  We then use the
Compton $y$ parameter, $y=0.25$ [$y\equiv 4(kT/m_{\rm e} c^2)\tau$, which
corresponds to $\Gamma\approx 1.9$] and the relative reflection strength
obtained from that fit in fitting our average spectrum of all Seyfert 1s
detected by OSSE. (Although the OSSE data alone only weakly constrain $y$,
$y=0.25$ is within $1\sigma$ of the best fit.)  This gives the optical depth
corresponding to the half-thickness of the slab of $\tau=0.33^{+0.07}_{-0.08}$,
which implies $kT=100_{-20}^{+30}$ keV (at $y=0.25$).  This spectrum is shown
in the left panel of Fig.\ \ref{sy_osse}.  We also consider another geometry,
of central hot sphere surrounded by a cold disk partly embedded in the sphere
(Poutanen, Krolik \& Ryde 1997).  For this geometry, the best-fit radial
optical depth of the central sphere is $\tau=1.3$.

Similar results are obtained for the average spectrum of Seyfert 2s (see the
right panel of Fig.\ \ref{sy_osse}), for which we also assume $y=0.25$ and the
relative strength of Compton reflection as for Seyfert 1s except that now
$i=60^\circ$.  Then the optical depth of the half-thickness of the slab is
$\tau=0.25^{+0.07}_{-0.06}$, corresponding to $kT=130_{-30}^{+30}$ keV and to
the radial optical depth of the central sphere of $\tau=1.1$.  We see that the
average plasma parameters of Seyfert 1s and 2s overlap within 90\% confidence,
consistent with the unified AGN model.

\section{Hard X-ray Power Laws and Compton Reflection} \label{s:refl}

Compton reflection (Lightman \& White 1988) is a process of backscattering of
X$\gamma$ radiation from a cold medium subtending some solid angle when viewed
from the primary source.  Compton reflection of a power-law spectrum produces a
characteristic spectral hump peaking at a few tens of keV, which is due to
bound-free absorption of the incident radiation below $\sim 10$ keV and the
inelasticity of Compton scattering above $\sim 100$ keV.  Compton reflection
can be measured by a relative normalization, $R$, of the {\it observed\/}
reflected component with respect to the primary one, where $R=1$ corresponds to
reflection of an isotropic source from an infinite slab.  If the primary source
is isotropic and neither the source nor the reflector are obscured, $R=
\Omega/2\pi$, where $\Omega$ is the solid angle subtended by the reflector as
viewed from the primary source.

Note that in many situations $R\neq \Omega/2\pi$.  First, the primary source
may be strongly obscured but a part of the reflector is directly visible, e.g.,
in Seyfert 2s, in which case we may have a reflection-dominated spectrum with
$R\gg 1$ (e.g.\ Reynolds et al.\ 1994).  Also, the primary source may be not
isotropic in the reflector frame (Ghisellini et al.\ 1990; Beloborodov 1999a,
hereafter B99a), which may either increase or decrease $R$ with respect to the
isotropic case.  Furthermore, the reflected component may be Compton-scattered
in the primary source (Poutanen \& Svensson 1996; Z98), which reduces the
observed $R$. These distinctions should be kept in mind when comparing an
observed strength of reflection, $R$, to a solid angle, $\Omega$, inferred in a
model.

Thus, $R$ is a quantity much more model-independent than $\Omega$.  A remaining
uncertainty in fitted values of $R$ is due to its dependence on the viewing
angle, $i$, of the reflecting plane, which is not known accurately for most
X$\gamma$ sources.  In general, the intensity of the reflected radiation
decreases with increasing $i$.  Furthermore, there is some dependence of
reflected spectra on geometry.  Below, we assume reflection as for an isotropic
irradiation of a slab (but with $R$ as a free parameter), for which case
viewing-angle dependent Green's functions are given by Magdziarz \& Zdziarski
(1995).

Continuum signatures of Compton reflection have been discovered in all of the
classes of sources considered here (Pounds et al.\ 1990; Nandra \& Pounds 1994;
Done et al.\ 1992; Ueda et al.\ 1994; Ebisawa et al.\ 1996; Gierli\'nski et
al.\ 1999; Tomsick et al.\ 1999).  Also Compton reflection has been found in
spectra of some weakly-magnetized neutron-star binaries (Yoshida et al.\ 1993;
Strickman et al.\ 1996) and white-dwarf binaries (e.g.\ Done, Osborne \&
Beardmore 1995; Done \& Osborne 1997; Done \& Magdziarz 1998; Cropper, Ramsey
\& Wu 1998).

\begin{figure}
\centerline{\psfig{file=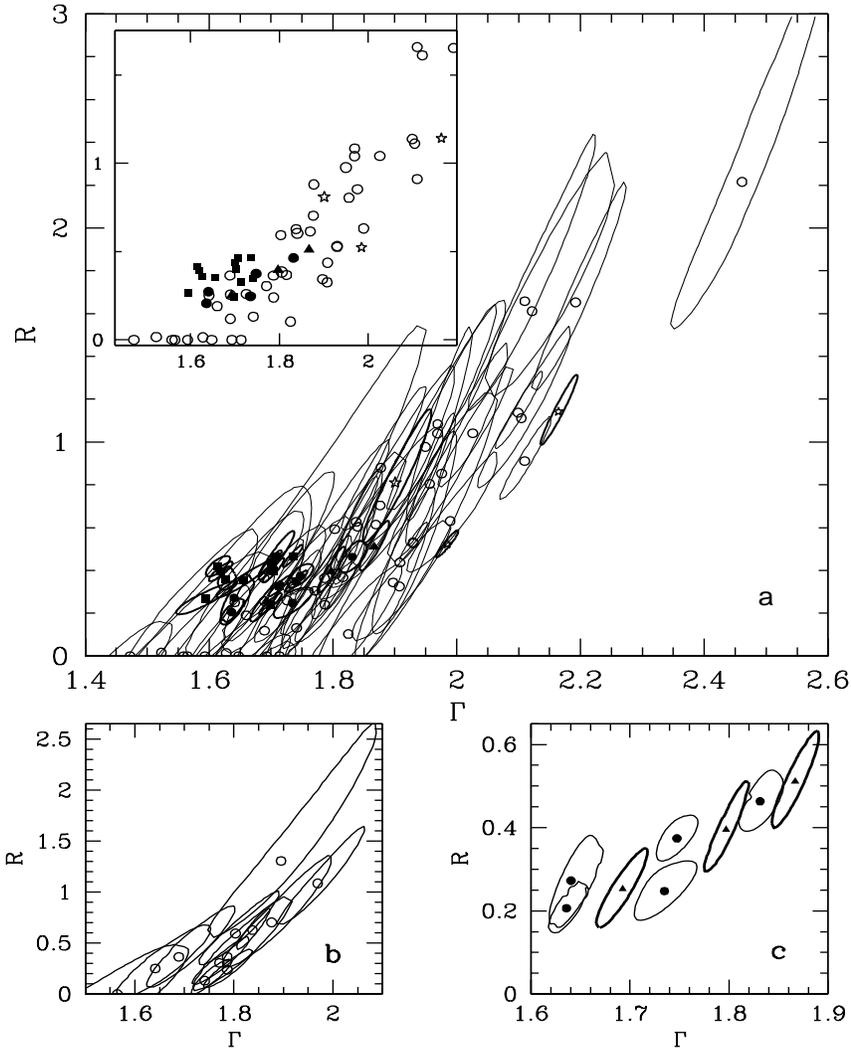,width=11cm,height=14cm}}
\caption{\small (a) The correlation of Compton
reflection with the intrinsic X-ray spectral index in Seyferts (thin contours)
and X-ray binaries in the hard state (heavy contours).  Open circles correspond
to Seyferts, and filled squares, circles, and triangles correspond to Cyg X-1,
GX 339--4, and Nova Muscae 1991, respectively. Asterisks correspond to 3
observations of 2 X-ray bursters. The inset shows the best-fit points without
error contours.  (b) Similar correlation occurring in multiple observations of
NGC 5548 and (c) GX 339--4 (thin contours and circles) and Nova Muscae 1991
(heavy contours and triangles).  } \label{correl} \end{figure}

A very strong correlation of the strength of Compton reflection with the
spectral index of the intrinsic spectrum, $\Gamma$, in both Seyferts and X-ray
binaries in the hard state has been found by Zdziarski, Lubi\'nski \& Smith
(1999, hereafter ZLS99).  Fig.\ \ref{correl} shows their results (with addition
of data for Nova Muscae 1991) for RQ Seyferts and X-ray binaries in the hard
state.  The observations used are by {\it Ginga}.  We see that all the contours
occupy a well-defined strip in the $(\Gamma, R)$ parameter space, with the
hardest spectra having almost no reflection and then with $R$ increasing with
$\Gamma$.  The softest Seyfert in the sample, MCG --6-30-15 (with $\Gamma\sim
2.5$) has somewhat less reflection that would be expected from extrapolation
from harder spectra.  Possibly, the reflection strength in Seyfert 1s saturates
at $R\sim 2$.

From Fig.\ \ref{correl}, we see that the error contours for individual spectra
typically span a much shorter range in the $(\Gamma, R)$ space than the extent
of the correlation itself.  Thus, the correlation cannot be an artifact of the
fitting method.  This is confirmed by a detailed statistical analysis performed
in ZLS99 using a method of Brandt (1997), which yields the significance of the
physical correlation alone (i.e., after removing the effect of the
measurement-related individual correlations) for the subsample of RQ Seyferts
of $>(1- 2\times 10^{-10})$. We note that spectra of Seyferts from {\it RXTE}\/
published so far (MCG --5-23-16, Weaver, Krolik \& Pier 1998; MCG --6-30-15,
Lee et al.\ 1998) show values of ($\Gamma$, $R$) within the area outlined by
the Seyfert confidence contours in Fig.\ \ref{correl}a, i.e., consistent with
the results of ZLS99.

As shown in ZLS99, correlations very similar to the global one are seen in
multiple observations of some individual objects.  Fig.\ \ref{correl}b shows
the correlation for NGC 5548, which was earlier found by Magdziarz et al.\
(1998).  The presence of this correlation represents a strong argument against
the global correlation being an orientation effect, as a change of the
orientation of an AGN on a short time scale (days in the case of NGC 5548) is
highly unlikely.

The contours for GX 339--4 (a black-hole candidate, see, e.g., Z98) are shown
in Figs.\ \ref{correl}a, c.  We see a very strong $R(\Gamma)$ correlation in 5
observations of this object (as found by Ueda et al.\ 1994). Also, Fig.\
\ref{correl}c shows 3 contours for {\it Ginga}\/ observations of Nova Muscae
1991 in the hard state (\.Zycki et al.\ 1998). We see an $R(\Gamma)$
correlation (as found by \.Zycki et al.\ 1999) indistinguishable from that in
GX 339--4.

As argued in ZLS99, the presence of the correlation implies a feedback in which
the presence of the reflecting cold medium affects the hardness of the X-ray
spectra.  A natural explanation for the feedback is that the cold medium emits
soft photons irradiating the X-ray source and serving as seeds for
thermal-Compton upscattering.  Then, the larger the solid angle subtended by
the reflector, the stronger the flux of soft photons, and, consequently, the
stronger cooling of the thermal plasma.  In a thermal plasma, the larger the
cooling rate, the softer the resulting X-ray power-law spectra. The evidence
that the X$\gamma$ spectra of both Seyferts and black-hole binaries in the hard
state are indeed formed by thermal Comptonization (Gierli\'nski et al.\ 1997;
Johnson et al.\ 1997; Grove et al.\ 1998; Zdziarski et al.\ 1996, Z97, Z98; see
\S\S 2, 3 above) strongly supports this explanation.  On the other hand, models
with X-ray emission being due to non-thermal electrons singly scattering seed
photons would yield no dependence of the X-ray slope on the cooling rate (e.g.\
Lightman \& Zdziarski 1987). Also, models with seed photons being mainly due to
a process intrinsic to the hot plasma, e.g.\ thermal synchrotron radiation,
would not reproduce the observed correlation.

\begin{figure}
\centerline{\psfig{file=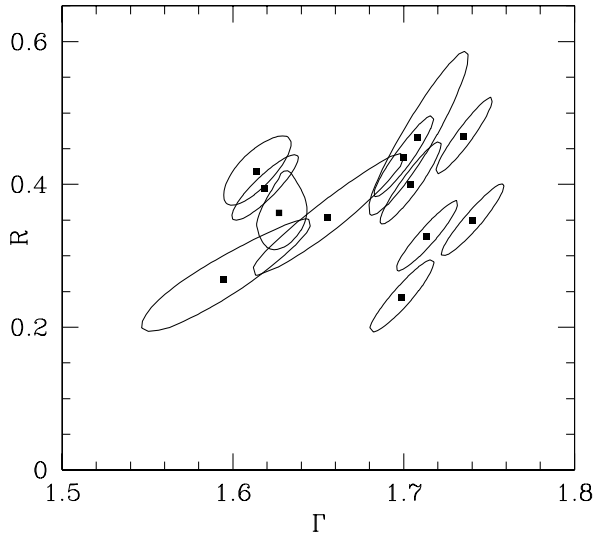,width=7.7cm,height=7.2cm}}
\caption{\small The strength of Compton
reflection in {\it Ginga}\/ observations of Cyg X-1 (assuming inclination of
the reflecting disk of $30^\circ$).}  \label{cygx1} \end{figure}

Thus, the presence of the correlation implies that both the X-ray emitting
plasma is thermal and that a substantial fraction of its cooling is due to soft
photons coming from the cold matter responsible for the observed Compton
reflection.  Within this general model, various specific geometries can in
principle reproduce the observational data presented in Fig.\ \ref{correl}.
This issue is discussed in \S 6 below.

In Fig.\ \ref{correl}a, we see that Compton reflection in Cyg X-1, although
with moderate $R$, is still significantly stronger than reflection for Seyferts
at the corresponding range of $\Gamma$.  The contours of the reflection
strength in the 12 {\it Ginga}\/ observations of Cyg X-1 in 1987, 1990 and 1991
are also shown separately in Fig.\ \ref{cygx1}.  We note that these contours
have been obtained assuming the source inclination of $30^\circ$.  If the
actual inclination is larger (Done \& \.Zycki 1999), the
values of $R$ implied by the data would further increase.  We also see in
Figs.\ \ref{correl}a, c that the strengths of Compton reflection in GX 339--4
and Nova Muscae 1991 are at least equal to the average $R$ of Seyferts.  Thus,
Compton reflection in black-hole binaries does {\it not\/} appear to be
anomously weak. Rather, reflection at a given $\Gamma$ is stronger on average
in black-hole binaries than in Seyferts.

This fact can be naturally explained in the framework of the model with thermal
Comptonization of blackbody photons emitted by the reflecting medium, e.g.\ an
optically-thick accretion disk.  The difference between sizes of stellar-mass
and supermassive black-hole sources implies different typical energies of
blackbody photons.  In inner disks around stellar-mass black holes, they are in
the soft X-ray range whereas they are in the UV range in the case of
supermassive black holes.  Sources with the same strength of reflection are
likely to have similar relative geometries of the hot plasma and the cold
medium, and thus similar ratios of the fluxes in the hard, Compton-upscattered,
radiation to those in soft, seed, photons incident on the hot plasma.  This
ratio is equal to the amplification factor, $A$, of the Comptonization process.

The X-ray spectral index, $\Gamma$, depends in general on $A$, in the sense
that the larger $A$, the harder the spectrum.  However, $\Gamma$ depends also
on the ratio of the plasma temperature, $T$, to the characteristic temperature
of the seed photons, $T_{\rm seed}$.  As discussed above, typical plasma
temperatures are similar in black-hole binaries and in Seyferts.  Then, at the
constant $T$, the larger $T_{\rm seed}$, the harder the X-ray spectrum (see
Beloborodov 1999b).  As seen in Fig.\ \ref{correl}a, this effect is seen in the
data.  For a given $R$, black-hole sources have smaller $\Gamma$ on average,
i.e., harder X-ray spectra.  In \S 6 below, the data are compared with
theoretical predictions for $R(\Gamma)$ at two values of $T_{\rm seed}$ for a
specific source geometry.

\section{X-ray and Soft Gamma-ray Spectra of Radio Galaxies} \label{s:radio}

As it is well established by now, X$\gamma$ emission of blazars is strongly
beamed and originates in their powerful jets (e.g.\ Sikora 1997).  The jets and
their lobes are also responsible for the strong radio emission of that class of
objects.  An interesting issue regards their nuclear, unbeamed, X$\gamma$
emission. In blazars, the X$\gamma$ emission of jets (oriented close to line of
sight) is so strong that it dominates completely any nuclear X$\gamma$
component.  To study the latter, we thus have to turn to radio-loud AGNs with
jets oriented at a significant angle with respect to the line of sight.

One suitable class of objects is nearby BLRGs.  X$\gamma$ emission of a sample
of them (3C 111, 3C 120, 3C 382, 3C 390.3, 3C 445) has recently been studied by
W98.  These authors used spectral data from {\it Ginga}, {\it ASCA}, {\it
EXOSAT}, and OSSE. Their main findings are summarized below.

\begin{figure}
\centerline{\psfig{file=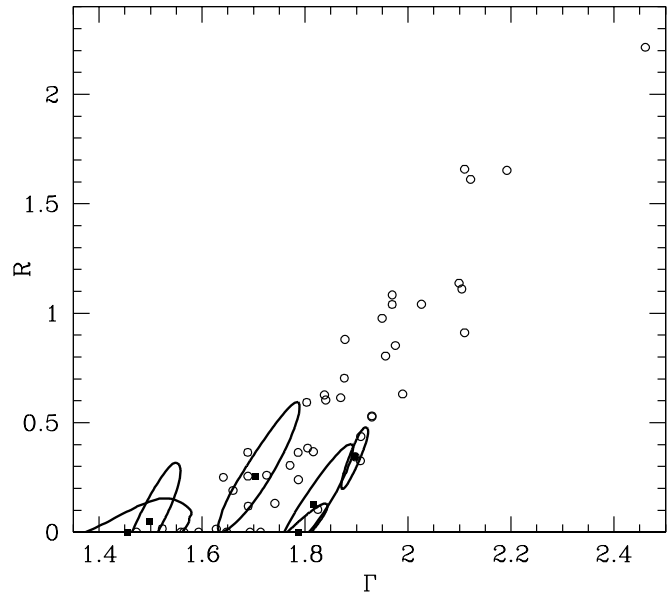,width=8.9cm}}
\caption{Compton reflection in radio galaxies
observed by {\it Ginga}\/ (fileed squares and contours) in comparison to that
in RQ Seyferts (open circles).  } \label{3c_fig} \end{figure}

First, the intrinsic X-ray spectra of BLRGs are harder and their reflection
components are weaker than those typical for RQ Seyferts.  For example, the
average {\it Ginga}\/ spectrum of 4 of the above BLRGs has $\Gamma=
1.67_{-0.04}^{+0.05}$, $R=0.08_{-0.08}^{+0.17}$, while we see in Fig.\
\ref{correl}a that Seyferts have $\Gamma\sim 2$, $R\sim 1$ on average.  On the
other hand, ZLS99 found that this difference does not imply that BLRGs and RQ
Seyferts belong to two different populations as far as their $\Gamma$ and $R$
are concerned.  Rather, BLRGs occupy a part of the $\Gamma$-$R$ parameter space
also populated by RQ Seyfert 1s.  This is shown in Fig.\ \ref{3c_fig}, which
shows the error contours for BLRGs and the nearby obscured radio galaxy Cen A
in comparison to the best-fit points for RQ Seyferts.

\begin{figure}
\centerline{\psfig{file=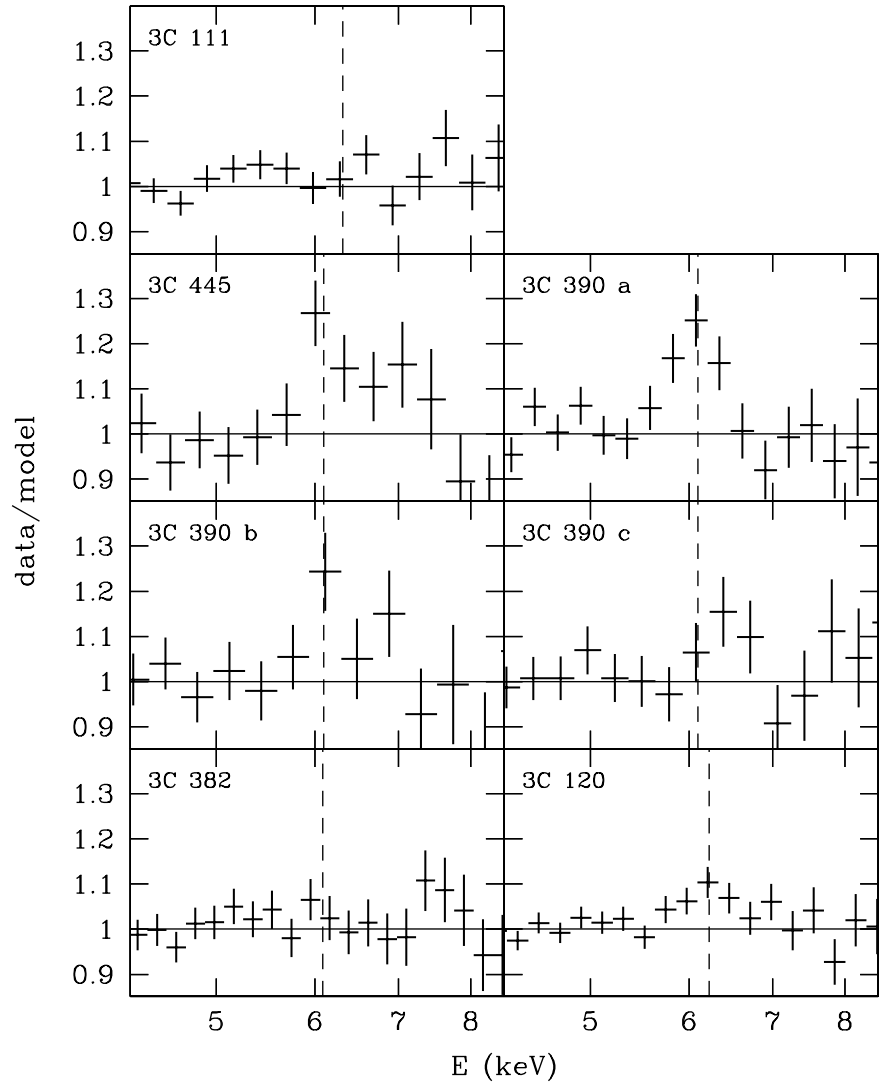,width=11.5cm}}
\caption{\small Fe K$\alpha$ line profiles of
BLRGs observed by {\it ASCA}, as analyzed by W98.  } \label{line_rl}
\end{figure}

Second, {\it ASCA}\/ observations of those objects show that their Fe K$\alpha$
lines are relatively weak and narrow.  The observed profiles are shown in Fig.\
\ref{line_rl}.  In all cases except 3C 120, the intrinsic line widths are
consistent with null within 90\% confidence (see Table 10 in W98).  In the case
of 3C 120 fitted with a broken power-law continuum without reflection and a
Gaussian line, both the line width, $\sigma_{\rm Fe} =0.32_{-0.20}^{+0.76}$
keV, and the equivalent width, $W_{\rm Fe} \simeq 100$ eV, are moderate.
Inclusion of reflection (not constrained by the {\it ASCA}\/ data) would reduce
the fitted $\sigma_{\rm Fe}$ and $W_{\rm Fe}$.

The Fe K$\alpha$ lines, although moderate, have still been found by W98 too
strong to be accounted for by the observed weak Compton reflection.  To solve
this problem, W98 proposed that a large fraction of the line flux originates in
a medium with the column density, $N_{\rm H}$, being $< 10^{24}$ cm$^{-2}$,
e.g.\ a molecular torus surrounding the nucleus.  The Thomson optical depth of
such a medium is then $<1$, and its Compton reflection does not give a hump
above $\sim 10$ keV (W98), characteristic to reflection from optically-thick
media. On the other hand, such a medium still can produce observable Fe
K$\alpha$ lines.  It can be estimated that for a solid angle of $\sim 2\pi$
subtended by that medium as seen from the central source, an $N_{\rm H} \sim
10^{23}$ cm$^{-2}$ is required to account for the observed line fluxes above
those due the observed Compton reflection (W98).  A medium with such $N_{\rm
H}$ is, in fact, observed in absorption in 3C 445 (as well as in Cen A).

We note that the results of W98 regarding the strength and the width of the Fe
K$\alpha$ lines in BLRGs differ from those of Reynolds (1997), Grandi et al.\
(1997) and Nandra et al.\ (1997), who generally found the lines to be stronger
and broader.  Some of the differences are accounted for by improvements in the
calibration of {\it ASCA}.  Those authors used its version of 1995 or earlier
whereas W98 used the release of 1997.  The effect of this change in the {\it
ASCA}\/ response and effective area can be illustrated by a comparison of fits
with the same model (power law plus a Gaussian line) to the same {\it ASCA}\/
spectrum of 3C 120 fitted in the 3--10 keV range.  For that spectrum, Nandra et
al.\ (1997) obtain $\sigma_{\rm Fe}= 0.74^{+0.34}_{-0.27}$ keV, $W_{\rm Fe}=
330^{+200}_{-120}$ eV, whereas W98 obtain $\sigma_{\rm Fe}=
0.28^{+0.96}_{-0.16}$ keV, $W_{\rm Fe}\approx 100$ eV.  Further, some of those
authors fitted the entire energy range of {\it ASCA}, $\sim 0.5$--10 keV, with
a single power-law continuum.  This neglects the soft excess, which was found
below $\sim 3$ keV in most observations of BLRGs in W98.  Given the limited
effective area of {\it ASCA}\/ above 7 keV, the claimed huge lines (e.g.\ with
$\sigma_{\rm Fe}\approx 2$ keV and $W_{\rm Fe} \approx 1$ keV in 3C 120 and 3C
382) are artifacts of compensating for the underestimated continuum at $\ga 6$
keV, with the underestimation due to fitting a single power law to a concave
intrinsic spectrum (W98).

The findings of W98 are consistent with results of Eracleous \& Halpern (1998)
for the BLRG Pictor A.  They find no detectable Fe K$\alpha$ line in the {\it
ASCA}\/ spectrum of that object, similarly to the case of 3C 111 (Fig.\
\ref{line_rl}). Although {\it ASCA}\/ data alone cannot determine the strength
of reflection, no detectable Fe K$\alpha$ line is compatible with $R\ll 1$ in
Pictor A.  Also, its power-law X-ray index, $\Gamma\simeq 1.76\pm 0.05$, is
consistent with the range of $\Gamma$ found in BLRGs in W98 (see Fig.\
\ref{3c_fig}).

\begin{figure}
\centerline{\psfig{file=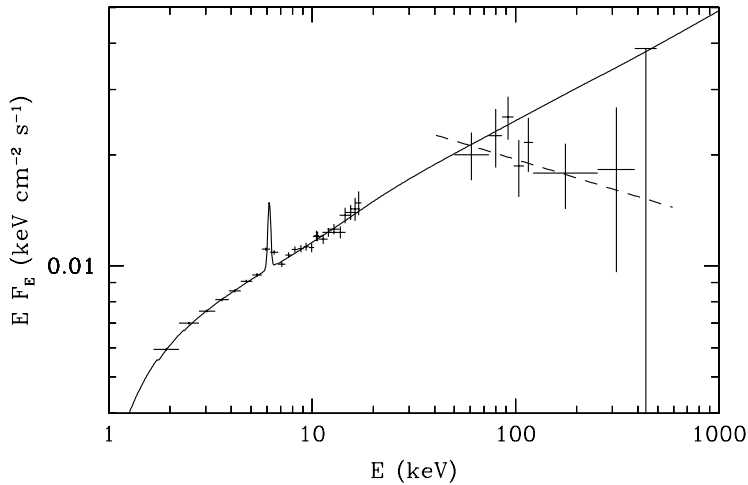,width=9.4cm}}
\caption{\small The average X$\gamma$ spectrum
of BLRGs from {\it Ginga}\/ and OSSE.  The solid curve gives the best fit of a
model with power law and (weak) Compton reflection to the {\it Ginga}\/ data,
and the dashed curve gives the power-law fit to the OSSE data.  }
\label{mean_rl} \end{figure}

W98 also studied the broad-band X$\gamma$ spectra of BLRGs.  Those spectra
break and become softer above $\sim 100$ keV, as shown by a simultaneous {\it
ASCA}/OSSE observation of 3C 120 and by the OSSE spectra being on average much
softer than the X-ray spectra.  Fig.\ \ref{mean_rl} shows the average X$\gamma$
spectrum of BLRGs from {\it Ginga}\/ and OSSE (W98).  The energy of the break
is similar to that seen in RQ Seyferts (\S\S 2, 3).  Also, the $(\Gamma, R)$
space occupied by BLRGs lies within that covered by RQ Seyferts, as shown in
Fig.\ \ref{3c_fig}.  The above facts support the origin of the X$\gamma$
spectra of BLRGs from thermal Comptonization, similarly as it seems to be the
case for RQ Seyferts.

On the other hand, the average OSSE spectrum of BLRGs does not show a
curvature, see Fig.\ \ref{mean_rl} (in contrast to that of RQ Seyferts), which
allows for an origin of X$\gamma$ photons from non-thermal Comptonization. This
conjecture appears to be supported by the form of the broad-band X$\gamma$
spectrum of Cen A.  In X-rays, $\Gamma\approx 1.7$--1.8 (W98; Steinle et al.\
1998) and $R=0^{+0.15}$ (W98), i.e., very similar to spectra of BLRGs.  The
spectrum breaks to a softer power law with $\Gamma\ga 2$ above 150 keV, and
then it breaks again above $\sim 10$ MeV, to $\Gamma\ga 3$, as required by an
EGRET detection of the source in the 0.1--1 GeV energy range (Steinle et al.\
1998). Observations of photons at such high energies indicate the presence of
non-thermal electrons in the X$\gamma$ source of Cen A.  The data on BLRGs are
of much more limited statistics and the issue whether their X$\gamma$ spectra
are from thermal Comptonization or a non-thermal process remains open so far.

\section{Geometry} \label{s:geo}

Results presented above show that the X$\gamma$ spectra of black-hole binaries
in the hard state are similar to those of Seyferts, and probably to those of
BLRGs. This suggests a source geometry common to those sources.  Within that
geometry, at least one variable parameter is required to account for the X-ray
hardness and the associated strength of Compton reflection varying from source
to source (as well as being variable in some individual objects).

Currently, there are two main competing pictures of the geometry of central
regions of black-hole binaries in the hard state and Seyferts.  In one picture,
the X$\gamma$ source forms a central hot disk (Shapiro, Lightman \& Eardley
1976). The plasma in that disk is two-temperature, with ions being much hotter
than the electrons, and with Coulomb energy transfer from the former to the
latter. The accretion-disk solution of Shapiro et al.\ (1976) is
cooling-dominated and it has been found thermally unstable by Pringle (1976).
However, there exists another solution, dominated by advection of hot ions into
the black hole, which seems to be stable (Abramowicz et al.\ 1995; Narayan \&
Yi 1995).  The intersection of the two solutions corresponds to the maximum
possible accretion rate and luminosity from the system.  As discussed, e.g., by
Zdziarski (1998), typical temperature and optical depth in the vicinity of that
maximum are $kT\sim 100$ keV, $\tau\sim 1$, which are just the average values
observed from Seyfert 1s and black-hole binaries in luminous hard states (see
\S\S 2, 3 above).

Observations of Compton reflection accompanying thermal Comptonization imply
the presence of optically-thick medium in the vicinity of the hot disk.  This
might be a standard optically-thick disk (Shakura \& Sunyaev 1973) surrounding
the hot disk.  In general, there may be an overlap between the two phases.

\begin{figure}
\begin{center}\leavevmode \epsfxsize 12.5cm\epsfbox{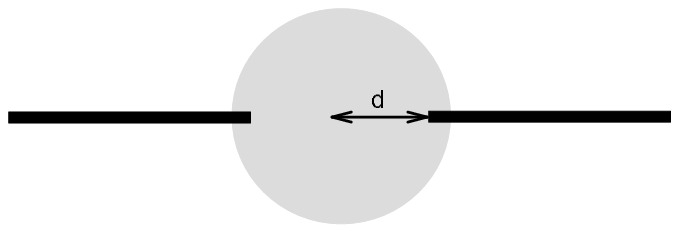}\end{center} 
\caption{\small The geometry with a central hot
plasma surrounded by a cold disk.  The cold disk enters partly the hot plasma
down to the radius $d$, which parameter determines the spectral hardness and
the amount of reflection present.  } \label{hot_disk} \end{figure}

\begin{figure}\begin{center}\leavevmode \epsfxsize 7.8cm
\epsfbox{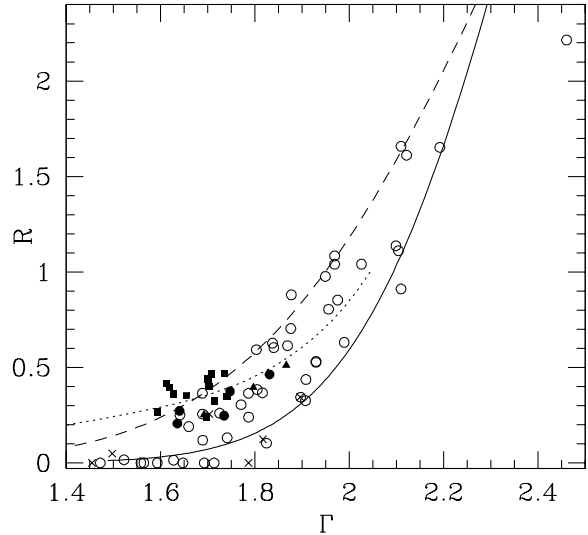}\end{center} \caption{\small The $R$-$\Gamma$ data together
with model dependences for an inner hot/outer cold disks (dotted curve), and
for mildly-relativistic bulk motion of the hot plasma above a cold disk in the
cases of seed photons at $kT_{\rm seed}=5$ eV and 200 eV (solid and dashed
curves, respectively).  Open circles and crosses correspond to Seyferts and
radio galaxies, respectively, and filled square, circles and triangles, to Cyg
X-1, GX 339--4 and Nova Muscae 1991, respectively.  } \label{models}
\end{figure}

We consider an idealized model of a central optically-thin sphere of unit
radius surrounded by a flat, cold, disk partly embedded in the hot sphere
(Fig.\ \ref{hot_disk}).  There is no intrinsic dissipation in the disk, the hot
sphere radiates via thermal Comptonization, and the seed photons for
upscattering are only the reprocessed ones emitted by the disk.  The relative
overlap of the hot and cold phases (as given by the radius $d$) determines then
the X-ray slope of the spectrum.  When $d\ga 1$, both cooling and the strength
of reflection are weak, and both increase with decreasing $d$.  This simple
model can be compared to the values of $\Gamma$ and $R$ observed in Seyferts
and hard-state black-hole binaries (ZLS99; \S 4), as shown by the dotted curve
in Fig.\ \ref{models} (for which $T_{\rm seed}=5$ eV was assumed).  We see that
this model reproduces the data for $R\leq 1$ reasonably well.

On the other hand, this simple model neglects the effect of Compton
upscattering of the reflected radiation in the hot plasma, which would then
reduce $R$ significantly, especially for $d\la 1$.  This effect has been taken
into account, e.g., in models of Esin et al.\ (1997, 1998) and Z98.  Their
general finding is that the strength of Compton reflection predicted in those
models is weaker than that observed in black-hole binaries.  The reflection
strength would increase in the presence of disk flaring.  Flaring is, in fact,
predicted by the standard model of Shakura \& Sunyaev (1973), and it is further
enhanced by irradiation (e.g.\ Vrtilek et al.\ 1990).  However, these effects
have been taken into account in the model of the hard state of Cyg X-1 by Esin
et al.\ (1998), who found the predicted reflection strength still less than
that observed.

This might be less of a problem for Seyferts with hard spectra, which often
have weak or null reflection, see Fig.\ \ref{models}.  On the other hand, there
are Seyferts with soft spectra and strong reflection, $R\sim 2$ (and strong
accompanying Fe K$\alpha$ lines), e.g.\ MCG --6-30-15.  If their geometry is
similar to that shown in Fig.\ \ref{hot_disk} for $d\ll 1$, $R$ has to be $<
1$.  In Seyferts, $R$ can be increased by both disk flaring and additional
reflection from Thomson-thick molecular torii surrounding the central source
(Ghisellini, Haardt \& Matt 1994; Krolik, Madau \& \.Zycki 1994).  However,
these effects should also be effective in some Seyferts with hard spectra,
contrary to the data in Fig.\ \ref{models} showing absence of Seyferts with
substantial reflection at $\Gamma\la 1.6$. Thus, the presence of reflection
with $R>1$ in some unobscured Seyferts is a problem for this model.  (If the
primary source is obscured but a part of the reflector is not, $R\gg 1$ can be
observed, e.g.\ Reynolds et al.\ 1994.)

Coming back to black-hole binaries, their models in terms of a hot inner disk
and an outer cold disk with a variable inner radius are supported by findings
of \.Zycki, Done \& Smith (1998) and Done \& \.Zycki (1999), who consider the
effect of relativistic smearing of both Compton reflection and the accompanying
Fe K$\alpha$ line.  The smearing occurs due to rotation of the disk in the
gravitational potential of the black hole (e.g.\ Fabian et al.\ 1989), and it
can be used to measure the range of radii where Compton reflection takes place.
\.Zycki et al.\ (1998) and Done \& \.Zycki (1999) find that the observed
reflection strength, $R$, roughly anticorrelates with the inner radius of the
cold disk (as inferred from relativistic smearing) in Nova Muscae and Cyg X-1.
This is consistent with the source geometry shown in Fig.\ \ref{hot_disk}.

The other geometry proposed for the X$\gamma$ sources in accreting black holes
is that of an active corona above the surface of an accretion disk. A likely
mechanism of energy dissipation in this case is magnetic flares (Galeev, Rosner
\& Vaiana 1979; B99a).  Haardt \& Maraschi (1993) proposed a model with a
homogeneous corona.  This model can work for soft sources provided the corona
optical depth is $\ll 1$, as to avoid strong attenuation of both reflection and
the Fe K$\alpha$ line due to scattering by hot electrons in the corona.  A
modified model with localized active regions (Haardt, Maraschi \& Ghisellini
1994) works well for objects with average $\Gamma$ and $R$ (Stern et al.\
1995).  However, it can be ruled out for black-hole binaries in the hard state
(Gierli\'nski et al.\ 1997; Z98).  The hard X-ray spectra of these objects
require the active regions to be elevated above the disk to reduce cooling,
which, however, results in $R\sim 1$, i.e., more than observed.

A solution to this problem has been proposed by B99a.  Namely, he postulates
that the active regions have some bulk motion, which can be due to the pressure
of reflected radiation and/or plasma ejection from magnetic flares.  A mildly
relativistic flow away from the disk, with a velocity $\beta c$, reduces then
the downward X$\gamma$ flux, which in turn reduces both reflection and
reprocessing. The reduction of the reprocessed flux incident on the hot plasma
leads to a spectral hardening (analogously to the previous model).  Two
$R(\Gamma)$ dependences of this model (for $i=30^\circ$ and for a
geometry-dependent parameter of B99a of $\mu_{\rm s}=0.4$) are shown in Fig.\
\ref{models}.  The solid and dashed curves correspond to seed photons at
$kT_{\rm seed}=5$ eV and 200 eV, characteristic to Seyferts and black-hole
binaries, respectively.  A given $R$ corresponds to the same $\beta$ on both
curves.  E.g., $R=0.3$, 0.5, 1, and 1.5 correspond to $\beta=0.27$, 0.15, 0,
and $-0.10$ (i.e., flow towards the disk), respectively.  We see that the
dashed curve reproduces quite well the data for black-hole binaries.  The solid
curve reproduces well the data for Seyferts with softest spectra for given $R$.
Then variations in the orientation, energy of the seed photons, and the plasma
parameters from source to source can reproduce the width of the correlation.  A
model best-fitting the Seyfert data is shown in Beloborodov (1999b) and ZLS99.
Comparing the solid and dashed curves, we also see that different values of
$T_{\rm seed}$ can indeed explain the black-hole binaries being harder on
average than Seyferts, as discussed at the end of \S 4.

Outflows were also proposed by W98 to explain the weakness of reflection in
radio galaxies.  We see that indeed the low-$T_{\rm seed}$ model of B99a
explains well those data (solid curve and crosses in Fig.\ \ref{models}).

Outflows cannot, of course, explain the data with $R>1$.  On the other hand, a
mildly relativistic motion directed towards the disk will enhance reflection
and cooling, leading to soft spectra with $R>1$, as also proposed by B99a.
Possibly, such downward motion can be due to ejection of plasma blobs from
magnetic flares towards the disk.  As seen in Fig.\ \ref{models}, the model of
B99a appears in principle capable to explain the entire observed correlation.

\section{Conclusions}

A general picture of X$\gamma$ sources in Seyferts and black-hole binaries in
the hard state that emerges from the results presented above is of a hot
thermal plasma with $kT\sim 10^2$ keV and $\tau\sim 1$ located in the vicinity
of a cold, optically-thick, medium.  The main radiative process in the hot
plasma is Compton upscattering of some soft seed photons.  The X-ray slope of
Compton-upscattered photons depends on the rate of cooling of the hot plasma by
the soft photons.

If the seed photons are emited by the reflecting medium, the strength of the
coupling determines the specific shape of emitted X$\gamma$ spectra.  When the
coupling between the hot and cold media is weak, there are few soft photons
incident on the hot plasma, which results in weak cooling and hard spectra.
Also, few hard photons from the hot plasma are incident on the cold medium and
Compton reflection is weak.  On the other hand, when the coupling is strong,
the plasma cooling is also strong, which yields soft X-ray slopes.  Also, many
hard photons are incident on the cold medium and reflection is strong.  This
explains the strong correlation between the intrinsic X-ray spectral index and
the relative strength of reflection found by ZLS99 (\S 4).

This general picture appears to be consistent with two geometries.  One is of a
hot inner disk surrounded by a cold outer disk.  The amount of overlap between
the hot and cold phases determines then the strength of the coupling (Fig.\
\ref{hot_disk}).  The other is of a hot plasma blob above the surface of a cold
disk moving perpendicular to the disk.  In the case of mildly relativistic
outflow, the coupling is weak, whereas motion towards to the disk leads to a
strong coupling.  Current data appear insufficient to conclusively determine
which of these geometries applies to observed black-hole sources.

Some differences between different classes of sources can be explained within
the above scenario.  Namely, the black-hole binaries have their intrinsic
spectra harder than Seyferts for given reflection strength, $R$.  This is
likely to be due the different characteristic blackbody energies in
stellar-mass systems and around supermassive black holes.  The higher blackbody
energies in black-hole binaries lead then to harder spectra for given
amplification of seed photons by thermal Comptonization (\S\S 4, 6).  The hard
spectra and weak reflection of BLRGs can be then connected to either
relativistic outflows of the hot plasma or due to the cold disks terminating
far away from the central black hole in those sources.

\acknowledgments

I thank Andrei Beloborodov for valuable discussions, and Ken Ebisawa and Piotr
\.Zycki for providing me with their {\it Ginga}\/ data on Cyg X-1 and Nova
Muscae 1991, respectively.  This research has been supported in part by the KBN
grants 2P03C00511p0(1,4) and 2P03D00614, and NASA grants and contracts.

\label{lastpage}

\end{document}